\documentclass[aps, prd, twocolumn,print, showpacs, nofootinbib]{revtex4}
\usepackage{amsmath}
\usepackage{amsfonts}
\usepackage{amssymb}
\usepackage{txfonts}
\usepackage{mathrsfs}
\usepackage{graphicx}
\usepackage{epsfig}
\usepackage{dcolumn}
\usepackage{bm}
\newcommand*{\prdfigscale}{0.85}
\newlength{\prdcolwidth}
\newlength{\figwidth}
\newlength{\doublewide}
\begin{document}

\title{Ultra-compact binary pulsars as continuous dual-line gravitational wave sources}
\author{Wen-Cong Chen$^{1,2\footnote{chenwc@pku.edu.cn}}$}

\affiliation{$^1$School of Science, Qingdao University of Technology, Qingdao 266525, China;
\\$^2$School of Physics and Electrical Information, Shangqiu Normal University, Shangqiu 476000, China;
}

\date{12 Apr. 2021}

\begin{abstract}
Binary millisecond pulsars (MSPs) are detached binary systems consisting of a MSP and a He white dwarf. If the initial orbital periods of binary MSPs are less than 0.3 day, they would evolve toward ultra-compact binary pulsars due to the rapid orbital shrinkage by the gravitational wave (GW) radiation. During the orbital decay, the MSP with an ellipticity would spin down by the GW radiation and the magnetic dipole radiation. Our calculations indicate that the angular momentum loss is dominated by the GW radiation when the ellipticities of the neutron stars (NSs) are in the range of $(1-50)\times 10^{-7}$, and the frequencies of high-frequency GW signals from the rotating NSs are $10-100$ Hz when the binary pulsars can be visible as low-frequency GW sources. These high-frequency GW signals are possible to be detected by the aLIGO and the third-generation GW detectors such as Einstein Telescope, depending on the frequencies and the distances. Therefore, some ultra-compact binary pulsars have an opportunity to become intriguing dual-line GW sources. By detecting the low-frequency GW signals, the NS mass can be accurately derived. A dual-line detection of two band GW signals could provide a constraint on the moment of inertia and the ellipticity of the NS. Thus the dual-line GW sources can be potentially applied to constrain the equation of state of the NS.
\end{abstract}

\maketitle

\section{Introduction}
The detection of gravitational wave (GW) event GW 150914 originated from the merger of double black holes marked the start of GW astronomy \citep{abbo16}, which provides a new channel to get the useful information from the remote objects. Especially, the merger event GW 170817 of double neutron stars (NSs) detected by the the Advanced LIGO and Advanced Virgo detectors announced the beginning of a new era of multimessenger astrophysics \citep{abbo17}. Comparing with catastrophic merger events of double compact objects, continuous GW signals could provide many valuable information on the evolution of the stars or the binary stars. Actually, another important scientific aim of GW detector such as the Advanced LIGO, Advanced Virgo,
and KAGRA is to search the continuous high-frequency GW signals from rapidly rotating NSs that is asymmetric with respect to their spin axis. However, it is very regretful that none continuous GW signal was detected so far.

In the early 2030s, the European Space Agency will launch a space GW detector Laser Interferometer Space Antenna
(LISA), in which the scientific aim is to detect the low-frequency GW signals in the Galaxy \citep{amar17}. The sensitive frequency band of the LISA are in the range of 0.1 mHz to 0.1Hz, which can be emitted from compact binary systems with orbital periods in the range of 20 s to 5 hours. According to the stellar and binary evolution theory, the possible LISA sources in the Galaxy were generally thought to include double white dwarfs (WDs) \citep{hils00,nele01a,nele01b,nele03,yu10,kapl12,krem17,lamb19,liu20}, NS-WD binaries \citep{taur18,yu21}, double NSs \citep{yu15,taur17}, AM CVn stars \citep{nele01c,nele04,nele03,liu21}, and ultra-compact X-ray binaries (UCXBs) \citep{chen20,chen21}. Recently, compact intermediate-mass black hole X-ray binaries were also proposed to be potential LISA sources \citep{chen20b}.

If a source can simultaneously emit both high-frequency and low-frequency GW signals, it will provide more valuable information on the orbital and spin evolution. Normally, a rapidly rotating millisecond pulsar (MSP) with an ellipticity can radiate continuous high-frequency GWs signals. If the MSP is in a compact binary system, the inspiral of the donor star would also produce continuous low-frequency GW signals. Therefore, compact binary MSPs were suggested to be potential dual-line GW sources, and can be used to constrain the equation of state of the MSP \citep{taur18}.

In the classical recycling model, the progenitors of binary MSPs are generally thought to be low-mass X-ray binaries (LMXBs) \citep{bhat91}. The NS accretes the material and angular momentum from the donor star by the Roche lobe overflow, and is spun up to a millisecond period \citep{alpa82}. Once the donor star evolves into a He WD after it depletes the hydrogen envelope, the system becomes a detached binary consisting of a MSP and a He WD, i. e. binary MSP \citep{liu11,shao12,taur18,chen20}. Subsequently, the binary MSP with an orbital period less than $\sim0.3~\rm days$ experiences a rapid orbital shrinkage due to pure GW radiation, and evolves toward low-frequency GW source that can be visible by the LISA \citep{taur18,chen20}. Meanwhile, the creation of a "mountain" on the surface of the new born MSP during the recycling stage leads to an ellipticity \citep{hask17}, which produces high-frequency GW signals, and causes the spin-down of the MSP by the GW radiation and magnetic dipole radiation. If the frequency of the GW yielding by the spin motion is still in the high-frequency band when the binary MSP appears as a low-frequency GW source, the binary MSP is indeed a dual-line GW source \citep{taur18}. If the binary MSP evolve into UCXB via the mass transfer triggering by the Roche-lobe overflow of the WD, it still have an opportunity to be visible as a dual-line GW source \citep{suvo21}.

\section{Low-frequency GW from binary MSPs}
The progenitor of a binary MSP is a binary system including a NS (with a mass of $M_{\rm ns}$) and a main-sequence donor star (with a mass of $M_{\rm d}$). After the donor star fills its Roche lobe by the nuclear evolution, it transfers the surface H-rich material onto the NS, and the system appears as a LMXB \citep{bhat91}. Once the donor star evolve into a stripped He core, it would decouple from the Roche lobe, and the LMXB becomes a detached binary MSP. Subsequently, the He core begins a contraction and cooling phase, the MSP will spin down by the GW radiation and the magnetic dipole radiation, and the orbit experiences a rapid shrinkage due to GW radiation.

The numerical calculations by the stellar evolution code show that the LMXB would evolve toward a binary MSP that can be visible by the LISA if the initial orbital period is less than the bifurcation period \citep{chen20}. Meanwhile, the He-WD masses in the binary MSPs appearing as LISA sources concentrate in a narrow range of $0.160-0170~M_{\odot}$ \citep{taur18,chen20}. There exist a good correlation between the WD mass and the orbital period \citep{savo87,rapp95,taur99}, hence the initial orbital period of the binary MSP is about $P_{0}=0.3$ day \citep{jia14}. The change rate of the orbital period due to GW radiation is
\begin{equation}
\dot{P}=-\frac{96G^{5/3}(4\pi^{2})^{4/3}}{5c^{5}}\frac{M_{\rm d}M_{\rm ns}}{(M_{\rm d}+M_{\rm ns})^{1/3}}P^{-5/3}=-KP^{-5/3},
\end{equation}
where $G$ is the gravitational constant, $c$ is the speed of light in
vacuo. Taking $M_{\rm ns}=1.6~M_{\odot}$ (we consider a mass growth of the accreting NS during the
recycled stage), and $M_{\rm d}=0.165~M_{\odot}$, it yields $K=8.0\times10^{-7}$. We then obtain the orbital period of the binary MSP as a function of the orbital decay time $t_{\rm de}$ as follows
\begin{equation}
P=\left(P_{0}^{8/3}-\frac{8Kt_{\rm de}}{3}\right)^{3/8}.\label{op}
\end{equation}

With the inspiral of the He WD, the binary MSP will emit continuous low-frequency GW signals if its orbital period is less than $1-2$ hours \citep{taur18,chen20}. The characteristic strain of low-frequency GW coming from the orbital motion for a four-years mission lifetime is \citep{evan87,taur18,chen20}
\begin{eqnarray}
h_{\rm low}\approx 3.75\times 10^{-19}\left(\frac{f_{\rm low}}{1~\rm mHz}\right)^{7/6}\left(\frac{\mathcal{M}}{1~M_{\odot}}\right)^{5/3}\nonumber\\
\times\left(\frac{1~\rm kpc}{d}\right),\label{hl}
\end{eqnarray}
where $d$ is the distance, $f_{\rm low}=2/P$ is the low-frequency GW frequency, and $\mathcal M$ is the chirp mass. Adopting $M_{\rm d}=0.165~M_{\odot}$, and $M_{\rm ns}=1.6~M_{\odot}$, the chirp mass for the detached binary MSP can be derived to be ${\mathcal M }=(M_{\rm d}M_{\rm ns})^{3/5}/(M_{\rm d}+M_{\rm ns})^{1/5}$$\approx0.40~M_{\odot}$. According to Eqs. (\ref{op})
and (\ref{hl}), we plot the evolutionary track of binary MSPs with a distance of 1 kpc and 10 kpc in the characteristic strain versus the GW frequency diagram (see also Figure 1). The initial GW frequency of binary MSPs appearing as the LISA sources are
0.33 and 0.72 mHz for a distance of 1 kpc and 10 kpc, respectively. It is worth noting that the orbital decay time is independent of the current GW frequency because $P_{0}$ is much larger than $P$. From Eq. (\ref{op}), the orbital decay time can be approximately estimated to be $t_{\rm de}=8.7~\rm Gyr$.

Because of the continuous shrinkage of the orbit, the He WD would fill its Roche lobe, and the binary MSP evolves into an UCXB. Utilizing a mass-radius relation of the donor star with a $n=3/2$ polytropic representation, and assuming that the donor star is
fully degenerate, and hydrogen depleted, the radius of the donor star is given by \citep{rapp87,prod15}
\begin{equation}
R_{\rm d}=0.0128 \left(\frac{M_{\odot}}{M_{\rm d}}\right)^{1/3}~\rm R_{\odot}. \label{rd}
\end{equation}
For $M_{\rm d}\leq0.8 M_{\rm ns}$, the Roche-lobe radius of the donor star is \citep{pacz71}
\begin{equation}
R_{\rm L}=0.462a\left(\frac{M_{\rm d}}{M_{\rm ns}+M_{\rm d}}\right)^{1/3}, \label{rl}
\end{equation}
where $a$ is the orbital separation.

At the UCXB stage, the donor star should fill its Roche lobe, so $R_{\rm d}=R_{\rm L}$. According to the Kepler's third law and Eqs. (\ref{rd}) and (\ref{rl}), the donor-star mass can be expressed as
\begin{equation}
M_{\rm d}=23.1\left(\frac{f_{\rm low}}{1~\rm Hz}\right)M_{\odot}. \label{md}
\end{equation}
Based on the narrow mass range of the He WD, the GW frequencies at the initial stage of UCXBs are derived to be in the range of
$6.9-7.4~\rm mHz$. According to Eq. (\ref{op}), the detection timescales of a binary MSP as LISA sources are estimated to be 180, and 23 Myr for a distance of 1 kpc, and 10 kpc, respectively. Therefore, the binary MSPs have an enough timescale to appear as LISA sources before they evolve into UCXBs. Certainly, these low-frequency GW signals can also be detected by
the space GW detectors including Taiji \citep{ruan20} and TianQin \citep{luo16,huan20}.

\begin{figure}
\begin{center}
\includegraphics[width=3.6 in, height = 2.8 in,
clip]{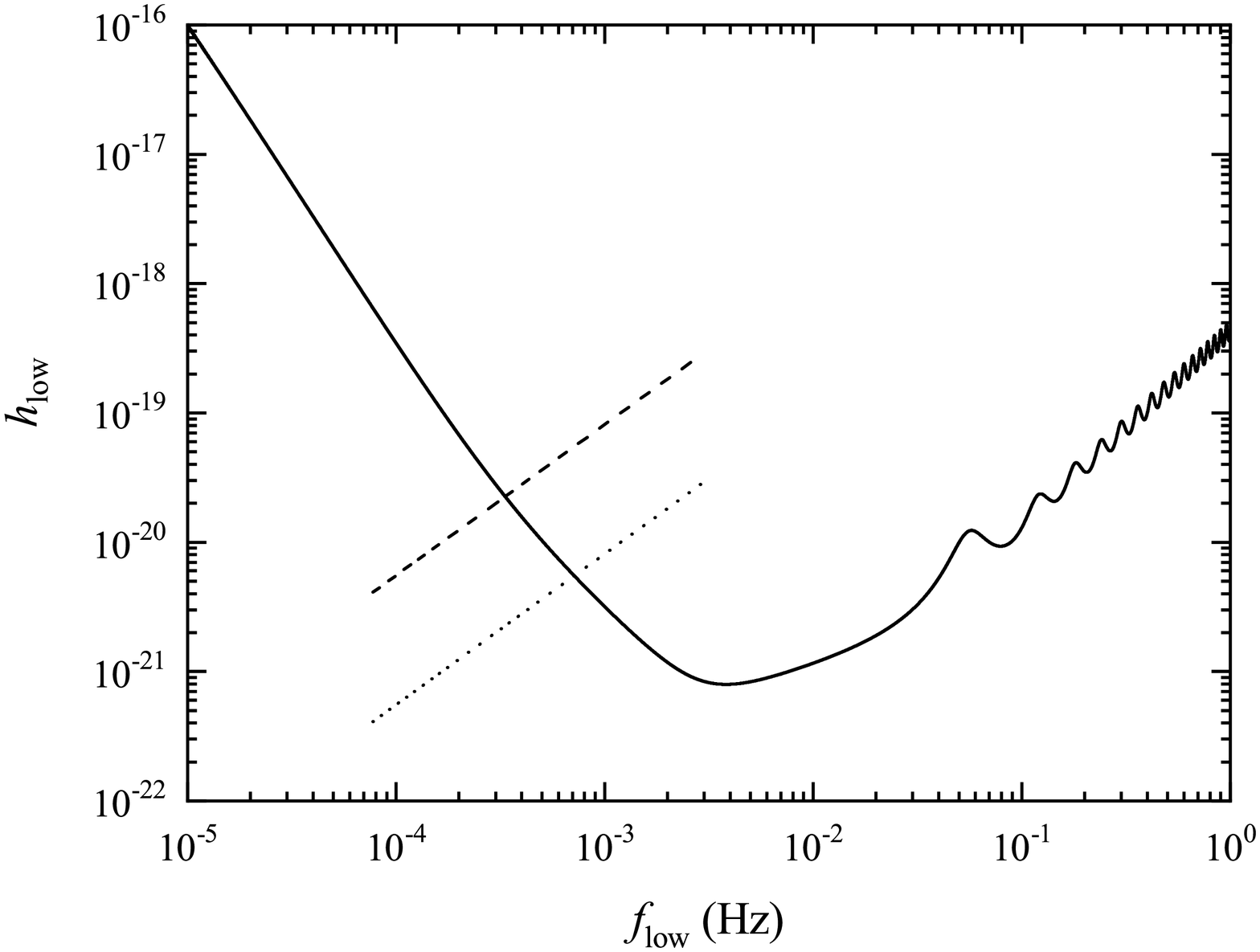} \caption{Evolution of the low-frequency GW signals producing by the orbital motion of binary MSP in the characteristic strain vs. the GW frequency diagram. The solid curve is the LISA sensitivity curve. The initial orbital periods of binary MSPs are assume to be 0.3 days. The dashed, and dotted curves corresponds to a distance $d=1$, and 10 kpc, respectively.} \label{Fig1}
\end{center}
\end{figure}

\section{High-frequency GW from a rotating NS}
A spinning NS that is asymmetric with respect to its spin axis can also radiate continuous GWs signals. In the LMXBs, the NS would be spun up to a millisecond period by accreting the material and angular momentum from the donor star \citep{alpa82}. Considering the effect of transient accretion in the accreting MSPs, the recycling channel could produce submillisecond pulsars \citep{bhat17,bhat21}, while they are not observed. Therefore, the MSP should experience a spin-down stage by the GW radiation and magnetic dipole radiation when the orbit of the binary MSP is rapidly shrinking (i.e. the LMXB evolves into a detached binary MSP) \citep{ande05}, and radiate the high frequency GW signals \footnote{The spin equilibrium produced by the accretion disk/magnetosphere interaction may account for the observed maximum spin frequency of MSPs \citep{patr12}.}. The existence of two subpopulations (with average spin frequencies of 300 Hz and 575 Hz) in accreting MSPs also hints the possibility of GW radiation extracting angular momentum \citep{patr17}.

The angular momentum loss rate of MSPs by the GW radiation is
\begin{eqnarray}
\dot{J}_{\rm gr}=-\frac{32\pi^{5}G\epsilon^{2} I^{2}f_{\rm high}^{5}}{5c^{5}}
=-5.4\times 10^{29}\nonumber\\
\epsilon_{-7}^{2}I_{45}^{2}\left(\frac{f_{\rm high}}{100~\rm Hz}\right)^{5}~\rm g\,cm^{2}s^{-2}.\label{gw}
\end{eqnarray}
where $\epsilon_{-7}=\epsilon/10^{-7}$, and $I_{45}=I/10^{45}$ are the ellipticity and the moment of inertia of the NS, $f_{\rm high}$ is the frequency of high-frequency GW signals. However, the maximum angular momentum loss rate of a MSP by the magnetic dipole radiation can be approximately estimated to be
\begin{eqnarray}
\dot{J}_{\rm md}=-\frac{2\pi^{3}B^{2}R^{6}f_{\rm high}^{3}}{3c^{3}}
=-7.7\times 10^{27}B_{8}^{2}R_{6}^{6}\nonumber\\
\left(\frac{f_{\rm high}}{100~\rm Hz}\right)^{3}~\rm g\,cm^{2}s^{-2},
\end{eqnarray}
where $B=B_{8}10^{8}~\rm G$, and $R=R_{6}10^{6}~\rm cm$ are the surface magnetic field and the radius of the NS.
Therefore, the angular momentum loss rate by the magnetic dipole radiation can be ignored for a MSP with typical parameters such as $\epsilon_{-7}=B_{8}=R_{6}=I_{45}=1$, and $f_{\rm high}=100~\rm Hz$.

If the GW radiation is the dominant mechanism influencing the spin evolution of the MSP, we have $I\pi\dot{f}_{\rm high}\approx \dot{J}_{\rm gr}$. Assuming that the ellipticity and the moment of inertia are constants,  the frequency of the high-frequency GW after a spin-down timescale $t_{\rm sd}$ is derived as
\begin{equation}
f_{\rm high}=\frac{f_{\rm high,0}}{(6.8\times 10^{-26}\epsilon_{-7}^{2}I_{45}f_{\rm high,0}^{4}t_{\rm sd}+1)^{1/4}} \label{fh}.
\end{equation}
Figure 2 shows the evolution of the frequency of the high-frequency GW signals. To detect the dual-line GW signal from a same binary MSP, it requires $t_{\rm sd}=t_{\rm de}$. In principle, $t_{\rm de}$ depends on the masses of two components and the initial orbital period of the binary MSP. However, the frequency of high-frequency GW signals is weakly dependent of $t_{\rm sd}$ according to Eq. (\ref{fh}). Taking $t_{\rm sd}=t_{\rm de}=8.7~\rm Gyr$, the NSs with three ellipticities expect for $\epsilon_{-7}=100$ emit GW signals with a frequency range of $\sim30-300~\rm Hz$. Therefore, the NS is still high-frequency GW sources when the NS+WD binary evolve into a compact binary pulsar that can be visible as a LISA source.

\begin{figure}
\begin{center}
\includegraphics[width=3.6 in, height = 2.8 in,
clip]{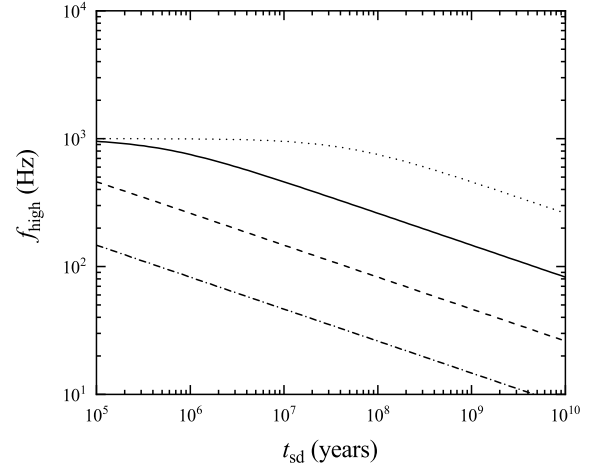} \caption{Evolution of the frequency of the high-frequency GW signals producing by the spin motion of MSPs with an ellipticity. In this figure, we take $f_{\rm high,0}=1000~\rm Hz$, and $I_{45}=1$. The solid, dashed, dotted, and dashed-dotted curves corresponds to an ellipticity $\epsilon_{-7}=1$, 10, 0.1, and 100, respectively.} \label{Fig1}
\end{center}
\end{figure}

The spin-down timescale of the GW radiation for a NS can be expressed as
\begin{equation}
\tau_{\rm gw}=-\frac{\pi If_{\rm high}}{\dot{J}_{\rm gw}}={18~\rm Gyr}\times \epsilon_{-7}^{-2}I_{45}^{-1}\left(\frac{100~\rm Hz}{f_{\rm high}}\right)^{4}  \label{time}.
\end{equation}
For a NS emitting GW signals with a frequency of 100 Hz, this timescale is obviously greater than the mission lifetime ($t_{\rm LISA}=4~\rm year$) of the LISA. During the LISA mission, the characteristic GW strain of a NS with an ellipticity $\epsilon$ can be approximately written as \citep{finn00,cors09}
\begin{equation}
h_{\rm high}=h_{0}f_{\rm high}\sqrt{\frac{dt}{df_{\rm high}}}\approx h_{0}\sqrt{f_{\rm high}t_{\rm LISA}} \label{hh},
\end{equation}
where the GW strain \citep{abbo07,hask15,abbo18}
\begin{equation}
h_{0}=\frac{4\pi^{2}G\epsilon If^{2}_{\rm high}}{c^{4}d} .
\end{equation}

Table I lists some input and derived parameters for NSs emitting high-frequency GW signals ($B_{8}=R_{6}=I_{45}=1$). When $\epsilon_{-7}=0.1$, and 0.01, the derived $f_{\rm high}$ is invalid because $\dot{J}_{\rm gr}$ is not much higher than $\dot{J}_{\rm md}$. However, the detection possibility of high-frequency GW from NSs with $\epsilon=10^{-9}-10^{-8}$ by the aLIGO and ET still remains because $f_{\rm high}$ is impossible to alter the orders of magnitude of $h_{\rm high}$. For ellipticities with $\epsilon_{-7}=1-50$, the frequencies of high-frequency GW signals are in the range of $12-94$ Hz. Figure 3 illustrates the distribution of samples in table I in the characteristic strain versus the GW frequency diagram. The NSs emitting high-frequency GW signals with $f_{\rm high}\sim100~\rm Hz$ within a distance of 1 kpc can be visible by the aLIGO, while the Einstein Telescope (ET) can detect a wide frequency range of $10-100~\rm Hz$ \citep{magg20}. For GW signals with a frequency of $\sim100~\rm Hz$, the detection horizon of the ET can reach $d=10~\rm kpc$. Actually, the magnetic field of MSPs may slightly exceed $10^{8}~\rm G$, while this only marginally decreases the frequency upper limit of high-frequency GW signals.

\begin{figure}
\begin{center}
\includegraphics[width=3.6 in, height = 2.8 in,
clip]{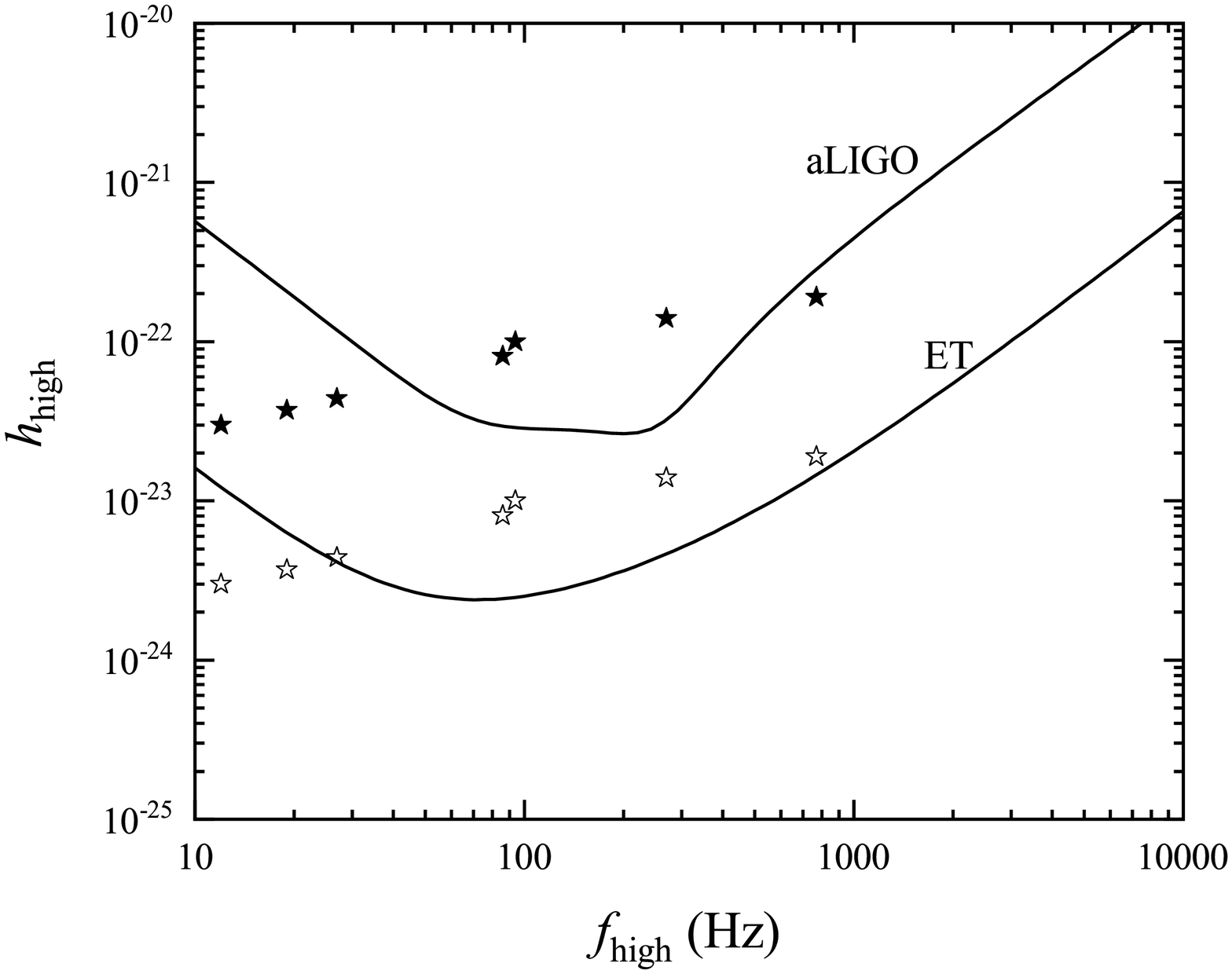} \caption{Distribution of some NS samples that can emitting high-frequency GW signals in the characteristic strain vs. the GW frequency diagram. The distances of the solid stars and the open stars are 1 kpc, and 10 kpc, respectively. The upper and bottom curves correspond to the sensitivity curves of the aLIGO and ET, respectively. The designed power spectral densities $S_{\rm n}(f)$ of aLIGO and ET arise from the analytical models (see also \citep{arun05,mish10,sun15}). } \label{Fig1}
\end{center}
\end{figure}

\begin{table*}
\begin{center}
\centering \caption{ Some input and derived parameters for NSs emitting high-frequency GW signals.\label{tbl-2}}
\begin{tabular}{ccccccccccc}
\hline\hline
Samples & initial spin period & $f_{\rm high,0} $& $d$&$t_{\rm de}$ & $\epsilon_{-7}$ &$f_{\rm high}$  &  $\dot{J}_{\rm gr}$& $\dot{J}_{\rm md}$& $\tau_{\rm gw}$ &  $h_{\rm high}$\\
\hline
 & (ms) & (Hz)   &(kpc) &(Gyr) &  & (Hz) &$(\rm g\,cm^{2} s^{-2})$ &$(\rm g\,cm^{2} s^{-2})$& (Gyr) &\\
\hline
1 & 1  & 2000   & 10  &  8.7  & 1 & 86  &$2.5\times10^{29}$& $4.9\times10^{27}$ &33  &$8.1\times10^{-24}$ \\
2 & 2  & 1000   & 10  &  8.7  & 1 & 86  &$2.5\times10^{29}$& $4.9\times10^{27}$ &33  &$8.1\times10^{-24}$ \\
3 & 2  & 1000   & 10  &  6.0  & 1 & 94  &$4.0\times10^{29}$& $6.4\times10^{27}$ &23  &$1.0\times10^{-23}$  \\
4 & 2  & 1000   & 10  &  8.7  &10 & 27  &$7.7\times10^{28}$& $1.5\times10^{26}$ &34  &$4.4\times10^{-24}$ \\
5 & 2  & 1000   & 10  &  8.7  &20 & 19  &$5.3\times10^{28}$& $5.3\times10^{25}$ &35  &$3.7\times10^{-24}$ \\
6 & 2  & 1000   & 10  &  8.7  &50 & 12  &$3.4\times10^{28}$& $1.3\times10^{25}$ &35  &$3.0\times10^{-24}$\\
7 & 2  & 1000   & 10  &  8.7  &0.1& 270 &$7.7\times10^{29}$& $1.5\times10^{29}$ &34  &$1.4\times10^{-23}$\\
8 & 2  & 1000   & 10  &  8.7  &0.01&769 &$1.5\times10^{30}$& $3.5\times10^{30}$ & 51 &$1.9\times10^{-23}$ \\
\hline\hline
\end{tabular}
 \end{center}
\end{table*}

\section{Constraining the equation of state of the NSs}
If a binary MSPs appears as a dual-line GW source, it has a possibility to
constrain the equation of state of the NSs. First, some LISA sources were expected to measure the frequency derivative $\dot{f}_{\rm low}$ \citep{amar12,shah12}.  By measuring $f_{\rm low}$, and $\dot{f}_{\rm low}$ of low-frequency GW signals, the chirp mass can be derived by \citep{taur18}
\begin{equation}
\mathcal{M}=\frac{c^{3}}{G}\left(\frac{5\pi^{-8/3}}{96}f^{-11/3}_{\rm low}\dot{f}_{\rm low}\right)^{3/5}.
\end{equation}
Subsequently, the NS mass can be accurately determined within $\sim4\%$ by the equation $\mathcal{M}=(M_{\rm ns}M_{\rm d})^{3/5}/(M_{\rm ns}+M_{\rm d})^{1/5}$ according to the narrow mass range of the WD. Obviously, it is difficult for UCXBs to accurately derive the NS mass by the chirp mass due to the contamination of the mass transfer.

Second, Eqs. (\ref{hl}) and (\ref{hh}) yield
\begin{eqnarray}
\epsilon I=1.0\times10^{44}\left(\frac{10~\rm Hz}{f_{\rm high}}\right)^{5/2}\left(\frac{f_{\rm low}}{1~\rm mHz}\right)^{7/6}
\nonumber\\
\times\left(\frac{\mathcal{M}}{1~M_{\odot}}\right)^{5/3}\left(\frac{h_{\rm high}}{h_{\rm low}}\right)~\rm g\,cm^{2} \label{ei1}.
\end{eqnarray}
It is worth noting that the moment of inertia is independent of the distance, and depends on the observation parameters $f_{\rm low}$, $f_{\rm high}$, $h_{\rm low}$, $h_{\rm high}$, and ${\mathcal M }$ apart from the ellipticity $\epsilon$.

Third, the angular momentum loss rate by the magnetic dipole radiation can be ignored if the frequencies of high-frequency GW signals are in a narrow range of $10-100$ Hz. From Eq. (\ref{gw}), we can obtain
\begin{equation}
\epsilon^{2} I=-5.8\times10^{34}\left(\frac{\dot{f}_{\rm high}}{10^{-17}~\rm Hz\,s^{-1}}\right)\left(\frac{10~\rm Hz}{f_{\rm high}}\right)^{5}~\rm g\,cm^{2}.\label{ei2}
\end{equation}
Meanwhile, a long term timing observations by the radio telescopes would yield the spin frequency $\nu$ ($f_{\rm high}=2\nu$) and the spin-frequency derivative $\dot{\nu}$ ($\dot{f}_{\rm high}=2\dot{\nu}$) of the NS in the binary pulsar, and confirm the source of high-frequency GW signals.  Combining Eqs. (\ref{ei1}) and (\ref{ei2}), the moment of inertia $I$ and the ellipticity $\epsilon$ can be derived. Meanwhile, the detection of low-frequency GW signals from the orbital motion presents an accurate determination on the NS mass. Therefore, the NS radius can be smoothly calculated according to the derived $I$ and $M_{\rm ns}$, and help understanding the unsettled equation of state (EOS) of NSs.

\section{Discussions}
According the above calculations, the maximum frequency of low-frequency GW signals is about $\sim 7$ mHz, over which the binary MSP has already evolved into UCXB. However, the minimum frequencies are in the range of 0.33 to 0.72 mHz (corresponding to orbital periods in the range 0.8 to 1.7 hours), which depend on the distance of the GW sources. Furthermore, the derived chirp mass is approximately $0.4~M_{\odot}$ due to a narrow range ($0.160-0170~M_{\odot}$) of WD masses \citep{taur18,chen20}. The frequency and chirp mass of low-frequency GW signals can be used to diagnose whether they originate from a binary pulsar.

Even if high-frequency and low-frequency GW signals are from a same sky region or with similar distances, we still can not conclude that the two band GW signals originate from a same source due to the rough determinations of the sky regions and the distances. The accurate confirmation has to employ the orbital periods and the spin periods of known binary pulsars. If the frequencies of two band GW signals are consistent with the observed parameters of a known binary pulsar, a dual-line GW source would be confirmed. Meanwhile, the comparisons of the sky regions and the distances between the two band GW signals and the known binary pulsar can help us to provide a constraint on the source of GW signals.

To accurately obtain the moment of inertia of the NS, the GW radiation should be dominant during the spin-down of a MSP.
Therefore, the frequencies of high-frequency GW signals concentrate in a narrow range of 10 to 100 Hz, which would be the detectable targets of the third-generation GW detector like ET. This frequency range is approximately consistent with that predicted by \citep{gao17}. Our predicted high frequency GW signals in dual-line GW can be radiated by NSs with
an ellipticity range of $(1-50)\times10^{-7}$. A strong internal toroidal magnetic field component ($B_{\rm t}$) could cause a NS distortion, producing an ellipticity $\epsilon=1.6\times10^{-6}(B_{\rm t}/10^{15}~\rm G)^{2}$ \citep{cutl02}. Therefore, the ellipticity of the new born millisecond magnetars can reach an order of magnitude of $10^{-3}$ \citep{gao16,mori16}. However, the MSPs forming in the recycling channel are NSs with weak magnetic field. The magnetohydrodynamic (MHD) simulation \citep{payne2004} show that the ellipticity of "magnetic mountain" created by the accreted matter can not exceed $\epsilon_{\rm MHD}\approx 2\times 10^{-7}$ \citep{payne2004,melatos2005,vigelius2009a}.  Coincidentally, the maximum elastic deformations of conventional NSs can only sustain a maximum elipticity of $2\times10^{-7}$ for a fiducial mass, radius, and breaking strain \citep{owen05}. However, since the shear modulus can be up to $4\times 10^{32}~\rm erg\,cm^{-3}$ \citep{xu03}, solid strange quark stars could possess a maximum ellipticity of $2\times10^{-4}$ \citep{owen05}. By the multimillion ion molecular dynamics simulations of Coulomb solids, the breaking strain of NS crust was proved to be around 0.1, which can support an ellipticity $\epsilon\leq4\times10^{-6}$ \citep{horo09}. Subsequently, the ellipticities of a canonical NS were proposed to be $\sim(1-8)\times10^{-6}$, which is dependent on the crust model and the EOS \citep{john13}. Therefore, the maximum ellipticity of a NS are still controversial in theory. The detections of dual-line GW sources provide us an opportunity to untie the mysterious veil of this issue.

Based on four known tightest binary MSPs containing a radio
MSP and a low-mass He WD \citep{manc05}, the number of Galactic NS+WD binaries that can be visible by the LISA is estimated to be $\sim 50$ \citep{taur18}. By modeling the observed binary pulsar population, the maximum number of ultra-compact NS-WD binaries in the Galaxy that are beaming towards the Earth was estimate to be $\sim1450$ \citep{pol20}. In theory, ultra-compact binary pulsars including NSs with an ellipticity of $\geq 10^{-9}$ are potential dual-line GW sources. Based on the distribution of spin periods and spin period derivatives of MSPs, the minimum ellipticity was constrained to be $\epsilon\approx10^{-9}$ \citep{woan18}. Recent estimations on the ellipticities of transitional MSPs and redbacks also focus at around $10^{-9}$ \citep{bhat20,chen20c}. Therefore, most of predicted 50 Galactic NS+WD binaries that can be visible by the LISA have opportunity to appear as dual-line GW sources. However, the constraint of EOS requires that the NSs have a high ellipticity of $(1-50)\times10^{-7}$. Very recently,  NSs with rotating period exceeding 12 ms and ellipticities of $\sim10^{-7}$ were not thought to be exist within a distance of 0.1 kpc \citep{stel20}. The NSs with a relatively high ellipticity should be very rare. According to the current theory on the ellipticity of NSs, it still remains possible to constrain the EOS of NSs by using the dual-line GW sources.

The pulsars in dual-line GW sources are braked by the torques producing by the GW radiation and magnetic dipole radiation. Therefore, the measured braking index of NSs should be in the range from 3 to 5 \citep{arch16,chen16}. To constrain the EOS of NSs, the ellipticity should be greater than $\geq 10^{-7}$. For such pulsars, the braking torques are virtually fully contributed by the GW radiation, hence the measured braking index should be approximately 5.

Certainly, it is impossible to detect the high-frequency GW signals for some ultra-compact binary pulsars due to a small ellipticity ($\epsilon<10^{-9}$) of NSs. In this case, the NS should be a MSP due to a small spin-down rate by the torque of weak GW radiation and magnetic dipole radiation. The measurement of low-frequency GW signals from the orbital motion can still provide an accurate constraint on the NS mass. However, the EOS of NSs still remains unsettled. If the binary pulsars evolve into UCXBs via the mass transfer triggering by the Roche-lobe overflow of WDs, thermonuclear X-ray burst of accreting NSs would provide some useful information in constraining their radius \citep{bhat05,bhat10}.

\acknowledgements  We thank the anonymous referee for
constructive comments that have led to the improvement of
the manuscript. We also thank Y. -W. Yu, R.-X. Xu, and E. P. Zou for helpful discussions. This work was partly supported by the National Natural Science Foundation of China (under grants no. 11573016, and no. 11733009).

\end{document}